\documentclass{article}
\usepackage{spconf,amsmath,graphicx}
\usepackage{amssymb,bbm}

\usepackage[breaklinks=true,letterpaper=true,bookmarks=false,hidelinks]{hyperref}
\usepackage{cleveref}
\usepackage{url}

\usepackage{graphicx}
\usepackage[caption=false]{subfig}

\usepackage{caption}

\usepackage{algorithm}
\usepackage{algorithmic}
\usepackage{booktabs}

\usepackage{array}
\newcolumntype{L}[1]{>{\raggedright\let\newline\\\arraybackslash\hspace{0pt}}m{#1}}
\newcolumntype{C}[1]{>{\centering\let\newline\\\arraybackslash\hspace{0pt}}m{#1}}
\newcolumntype{R}[1]{>{\raggedleft\let\newline\\\arraybackslash\hspace{0pt}}m{#1}}

\title{Optimize what matters: Training DNN-HMM Keyword Spotting Model Using End Metric}
\name{Ashish Shrivastava \qquad Arnav Kundu \qquad Chandra Dhir \qquad Devang Naik \qquad Oncel Tuzel
\thanks{Emails:  \{ashish.s, a\_kundu, cdhir, naik.d, otuzel\}@apple.com}
\vspace{-0.15in}
}
\address{
Apple }
\graphicspath{{./Figures/}}

\begin{document}
%
\maketitle
\begin{abstract}
Deep Neural Network--Hidden Markov Model (DNN-HMM) based methods have been successfully used for many always-on keyword spotting algorithms that detect a wake word to trigger a device.
The DNN predicts the state probabilities of a given speech frame, while HMM  decoder combines the DNN predictions of multiple speech frames to compute the keyword detection score.
The DNN, in prior methods, is trained independent of the HMM parameters to minimize the cross-entropy loss between the predicted and the ground-truth state probabilities.
The mis-match between the DNN training loss (cross-entropy) and the end metric (detection score) is the main source of sub-optimal performance for the keyword spotting task.
We address this loss-metric mismatch with a novel end-to-end training strategy that learns the DNN parameters by optimizing for the detection score.
To this end, we make the HMM decoder (dynamic programming) differentiable and back-propagate through it to maximize the score for the keyword and minimize the scores for non-keyword speech segments.
Our method does not require any change in the model architecture or the inference framework; therefore, there is no overhead in run-time memory or compute requirements.
Moreover, we show significant reduction in false rejection rate (FRR) at the same false trigger experience ($> 70\%$ over independent DNN training).

%
\end{abstract}

\section{Introduction}
Common strategies to improve a machine learning model are adding more training data, or using a model with more parameters/better architecture.
However, adding more data is not helpful with small footprint models with limited capacity (such as for keyword spotting), and adding more parameters increases runtime memory or compute, which is not possible for many real-time production systems.
We show that, without adding more data or changing the model architecture, we can significantly improve the performance by designing an optimization procedure that focuses on the end metric, particularly for low capacity models.

We focus on speech based natural interaction, which requires a device to constantly listen to and decode the streaming audio.
For power efficiency purposes, the device usually relies on a voice trigger interface that triggers a more compute-intensive pipeline if a specific keyword is detected.
A keyword spotting (KWS) algorithm, with small memory and compute footprint, is an essential component of the system that computes a detection score for every speech frame (typically $10$ms apart) and triggers if the score exceeds a threshold.

In DNN-HMM based KWS models, the DNN computes the state probabilities and the HMM decoder combines these probabilities using dynamic programming (DP).
We follow the same architecture as~\cite{SigtiaHRMB18} which contains $20$ states -- $18$ corresponding to $6$ phonemes of the trigger phrase ($3$ states for each phoneme), $1$ state for silence, and $1$ state for background.
The DNN uses a softmax layer to output probability distribution over $20$ classes corresponding to these $20$ states for each speech frame, and is trained to minimize the average (over all frames) cross-entropy loss between the predicted and the ground-truth distributions.
This training ignores the HMM transition and prior probabilities which are learned independently using training data statistics.
Such an independently trained DNN model relies on the accuracy of the ground-truth phoneme labels as well as the HMM model.
This model also assumes that the set of keyword states are optimal and each state is equally important for the keyword detection task.
The DNN spends all of its capacity focusing equally on all of the states, without considering its impact on the final metric of the detection score, resulting in a loss-metric mismatch.

To address this loss-metric mismatch, we train the DNN model by directly optimizing the keyword detection score instead of optimizing for the state probabilities.
This end metric based training uses only the start and the end of the keyword instead of requiring all of the speech frames to be annotated, leading to substantial savings in annotation cost.
Our method changes only the training algorithm without changing any inference pipeline; therefore, there is no overhead in runtime memory or compute, since we only need to update the model parameters.
We design an optimization procedure that maximizes the detection score for a speech segment that “tightly” contains the keyword (we call these positive samples) and minimize the detection score for the speech that does not contain the keyword (negative samples).
We also sample additional hard negatives that contain partial keywords because we do not want the model to trigger at partial phrases. 
To formalize the concept of “tightly” containing the keyword, let a ground-truth window start at time $g_1$ and end at $g_2$, and a sampled window (i.e. a set of contiguous speech frames) start and end at time $w_1$ and $w_2$, respectively.
We sample positive and negative windows from speech utterances such that the positive windows have high intersection-over-union ($IOU$) and negative windows have low $IOU$ with the ground-truth keyword window.
The $IOU$ between the ground-truth window $[g_1, g_2]$ and a sampled window $[w_1, w_2]$  is defined as:
\begin{align}
IOU &= \frac{\text{area of intersection}}{\text{area of union}} \nonumber \\
&= \frac{\max(0, \min(g_2, w_2) - \max(g_1, w_1))}{\max(g_2, w_2) - \min(g_1, w_1)}.
\end{align}
The $IOU$ based sampling in our method is inspired by image detection algorithms such as~\cite{yoloRedmon2016, yolo3Redmon2017, FasterRcnn2015}.
We differentiate through the DP algorithm of the  HMM decoder to optimize the detection scores of positive and negative windows.

Our main contributions include: (1)~defining the optimization of KWS model using the end metric which triggers only on a tightly overlapping (based on $IOU$) window with the ground-truth keyword,
(2)~learning the DNN parameters by computing gradients of this loss function through the HMM decoder, (3)~reducing the FRR by $>70\%$ (at the same FA) without adding more data or changing model architecture.

\section{Related Works}
HMM based architectures have been common in keyword detection \cite{mi_hmm_Bahl_1986, rohlichek1989,rose1990,wilpon1990automatic,alphanet_Bridle90}, where an acoustic model
computes the observation probabilities and the HMM decoder computes the detection score using the observation, the state transition, and the prior probabilities.
The DNN-HMM based models~\cite{SigtiaHRMB18,google2014,Chen2013AHH} compute the observation probabilities using a DNN with a softmax layer, which is trained separately to minimize the cross-entropy loss between the predicted and ground-truth observation probabilities.
Such optimization results in a loss-metric mismatch, where the metric (i.e. the detection score) is not directly optimized.
In contrast, our method optimizes the detection score explicitly.

Discriminative training methods have been proposed for speech recognition (including keyword spotting) tasks~\cite{dis_KWS_keshet_2009,seq_dis_KWS_keshet_2018} for generative models such as HMM.
We also propose a discriminative training framework; however, we specifically optimize the detection score based on positive and negative windows defined by $IOU$ criteria (inspired by object detection methods in images~\cite{yoloRedmon2016, yolo3Redmon2017, FasterRcnn2015}) and learn DNN parameters by differentiating through the HMM decoding.
Recurrent neural networks (RNNs) have also been used for keyword spotting~\cite{fernandez2007application,woellmer2013keyword,lengerich2016end,hwang2015online,graves2006connectionist}, but are generally not suitable for low-power streaming audio applications.
Recently, convolutional neural networks (CNN) based end-to-end trained models \cite{Zhang2017HelloEK,Tang2018DeepRL,google2015} have been shown to work better than the HMM based model.
However, these models require more computation to encode a large context.
Researchers have optimized the CNN based models~\cite{svdf2018,svdf2015,Sun2017CompressedTD,Higuchi2020}, but they minimize the frame level cross-entropy loss and suffer from loss-metric mismatch.
Compare to these CNN based models, the proposed model is more interpretable (predicts state probabilities), has a larger receptive field, better localization, and addresses loss-metric mismatch.

\section{The Proposed End-to-End Optimization}
\begin{figure}[t]
    \centering
    \includegraphics[width=.92\linewidth]{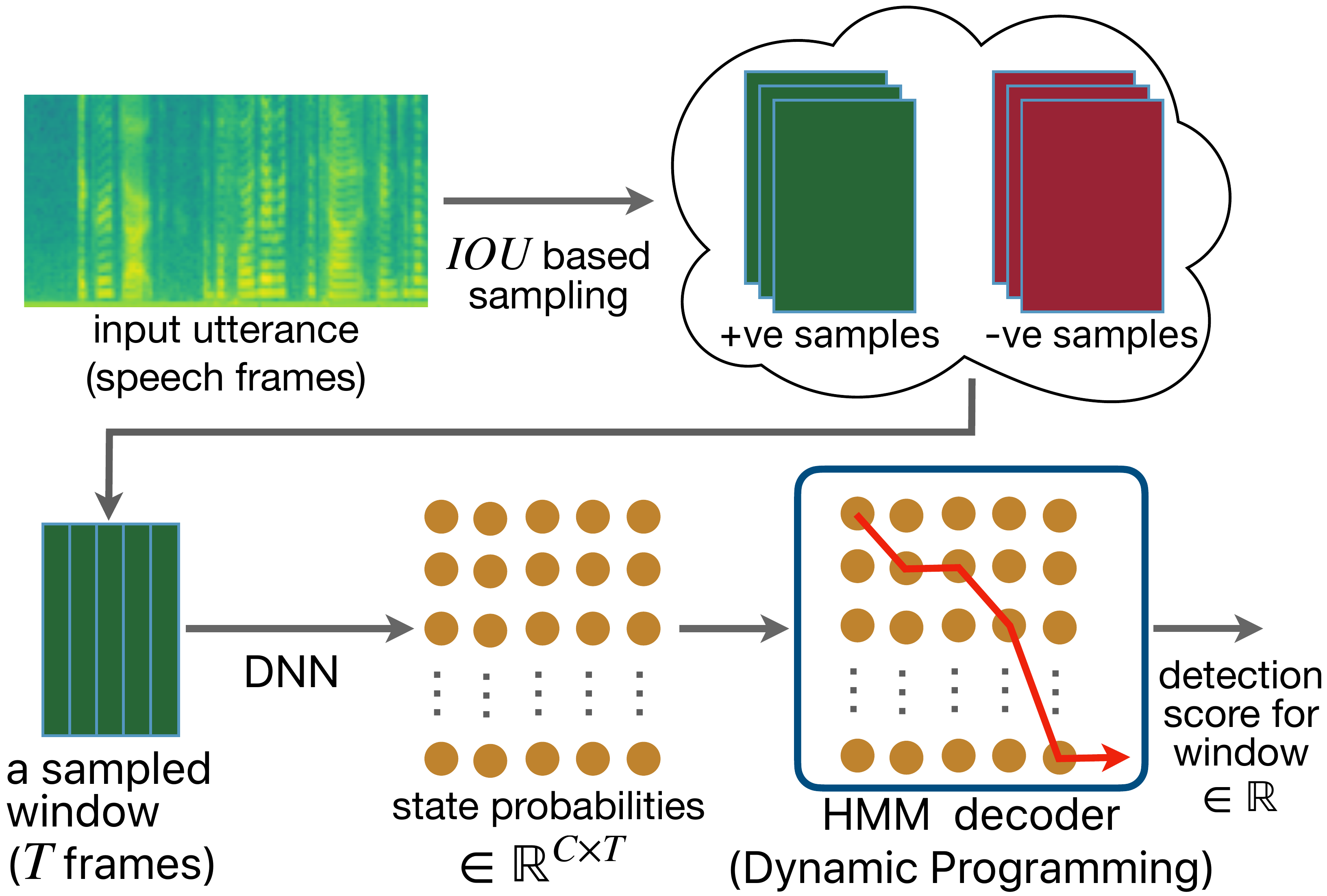} \\
    \vspace{-0.1in}
    \caption{Overview of the proposed optimization. 
    Positive samples have large $IOU$ and negative samples have a small $IOU$ with the  ground-truth keyword window.
    The detection score of the window corresponds to the most likely HMM path that starts at the beginning of the window and ends at the last frame of the window.
    We make the HMM differentiable and then update DNN parameters so that the scores of positive samples are high and the scores for negative samples are low.}
    \label{fig:method_overview}
\end{figure}

Each speech frame is labeled with one of the $C$ states which correspond to keyword phonemes, silence, and background speech. 
The DNN, consisting of a few fully-connected layers followed by a softmax layer, takes context dependent speech features, $\boldsymbol x_t \in \mathbb R^{d}$, at time $t$ and maps it to the probability distribution over $C$ states.
Following the architecture of~\cite{SigtiaHRMB18}, we use $13$-dimensional MFCC features~\cite{Mermelstein1976DistanceMF, Davis80comparisonof} computed over small overlapping windows ($25$ms in duration and separated by $10$ms) of the input audio.
To model context, the input feature $\boldsymbol x_t$ is generated by concatenating the MFCC features over a small window of frames $[t-\delta, t+\delta]$, where $\delta$ is a small number ($9$ in our implementation).

Let the DNN be denoted by $f_{\boldsymbol \theta}: \mathbb R^d \rightarrow \mathbb R^C$, where $\boldsymbol \theta$ are the DNN parameters.
Given a set of speech features $\boldsymbol x_j$
and their corresponding ground-truth state-labels, $y_j$, the DNN, in conventional methods, is independently trained by minimizing the following cross-entropy loss:
\begin{align}
\mathcal L_{ce} = \min_{\boldsymbol \theta} \sum_{j} -\log(f_{\boldsymbol \theta}(\boldsymbol x_j)[y_j]).
\label{eq:loss_ce}
\end{align}
During inference, the HMM decoder combines the DNN predictions using DP to compute the detection score of the keyword.
It computes the most likely path (and its probability) that starts at the state corresponding to the beginning of the keyword and ends at time $t$ at the state corresponding to the end of the keyword.
The HMM decoder backtracks the most likely path from the end of the keyword at time $t=T$, and finds the time $t_s$ for the beginning of the keyword on this most likely path.
The detection score, at time $t$, is computed as the probability of the most likely path ending at the state corresponding to the end of the keyword and is normalized by detection window size ($t - t_s$).
In conventional DNN-HMM models, only the DNN is trained and HMM is used only during inference.

In the proposed method (Figure~\ref{fig:method_overview}), positive windows are sampled such that their $IOU$s with the ground-truth keyword window is greater than a threshold, $iou_p$, and the negative windows are sampled to have their $IOU$s less than a threshold, $iou_n$.
Furthermore, we augment the negative training data by dividing the ground-truth window into two parts and swapping their order.
Such perturbation of the ground-truth window creates a hard negative sample and enforces the model to predict a high detection score only if the states are observed in the right order.
Intuitively, this helps the model to reject the incorrect ordering of the words in the trigger phrase.

We train the DNN model such that the positive windows have high scores and the negative windows have low scores.
To compute the detection score of a window, we use both the DNN and the HMM models during training.
We assume the start and the end of the windows are the start and the end of the trigger phrase.
The HMM decoder combines the DNN outputs of an input window to compute the most likely path that starts at the first frame and ends at the last ($T^{\text{th}}$) frame of the window.

At each time step, the HMM decoder computes the maximum probability of reaching the state $s_i, \forall i = 1, \dots, C$.
Let this probability be $v_i(t)$ at frame $t$, which can be written as:
\begin{align}
v_i(t) &=\max_{\boldsymbol s_1, \dots \boldsymbol s_{t-1}} Pr(\boldsymbol s_1, \dots, \boldsymbol s_{t-1}, \boldsymbol x_1, \dots, \boldsymbol x_t, \boldsymbol s_t = i), \nonumber \\
&= \max_{\boldsymbol s_1, \dots \boldsymbol s_{t-1}} \{v_i(t-1)*b_{i, i}, v_{i-1}(t-1)*b_{i-1, i}\} \nonumber \\
  & \quad \quad \quad \quad \quad \quad * Pr(\boldsymbol x_t | \boldsymbol s_t=i),
\end{align}
where $b_{i,j}$ are the transition probabilities of the HMM.
The observation probability given the model in state $i$, $Pr(\boldsymbol x_t | \boldsymbol s_t=i)$, is proportional to the DNN output for the $i^{\text{th}}$ class, $f_{\boldsymbol \theta}(\boldsymbol x_t)[i]$.
First, $v_i(0)$ is initialized with the prior probabilities and then iteratively updated using a max-pool layer by combining the transition probabilities and the DNN predictions.
After $T$ updates, the detection score of the window is $d = v_C(T) / T$, where division by $T$ is used to normalize the detection scores to account for different sequence lengths.
We use the hinge loss~\cite{hinge_loss}, which ignores the samples from optimization if their scores are beyond a margin:
\begin{align}
  \mathcal L_{e2e} = \min_{\boldsymbol \theta} \sum_{j \in \mathcal {\boldsymbol X}_p} \max(0, 1 - d_j) + \sum_{j \in \mathcal {\boldsymbol X}_n} \max(0, 1 + d_j),
  \vspace{-0.2in}
  \label{eq:loss_e2e}
\end{align}
where $\mathcal {\boldsymbol X}_p$ is the set of all the positive windows, $\mathcal {\boldsymbol X}_n$ is the set of all the negative windows, and $d_j$ is detection score for the $j^{\text{th}}$ window.
Note that this loss function is piece-wise smooth and can be optimized using gradient descent.
%

%
\section{Experiments}
 
\subsection{Data}

Similar to \cite{Higuchi2020}, our training data contains approximately $500k$ speech utterances, each containing one trigger phrase.
We apply simple augmentations on the data such as gain augmentation (multiplying the raw audio by a scalar) and room impulse response (RIR).
Our test data consists of approximately $2000$ hours of ``dense" speech (i.e. with little silence data) to test the robustness of the model.
The positive test data (i.e. utterances containing the trigger phrase) includes near field samples from a device held in the speaker's hand and far field samples from devices roughly $3$ft away from the speaker.
This data has been collected via internal user studies with informed consents.
The ground-truth keyword window is determined using the state labels corresponding to the beginning and the end of the keyword.

\begin{figure}[h!]
  \centering
\subfloat{%
  \includegraphics[clip,width=0.8\columnwidth]{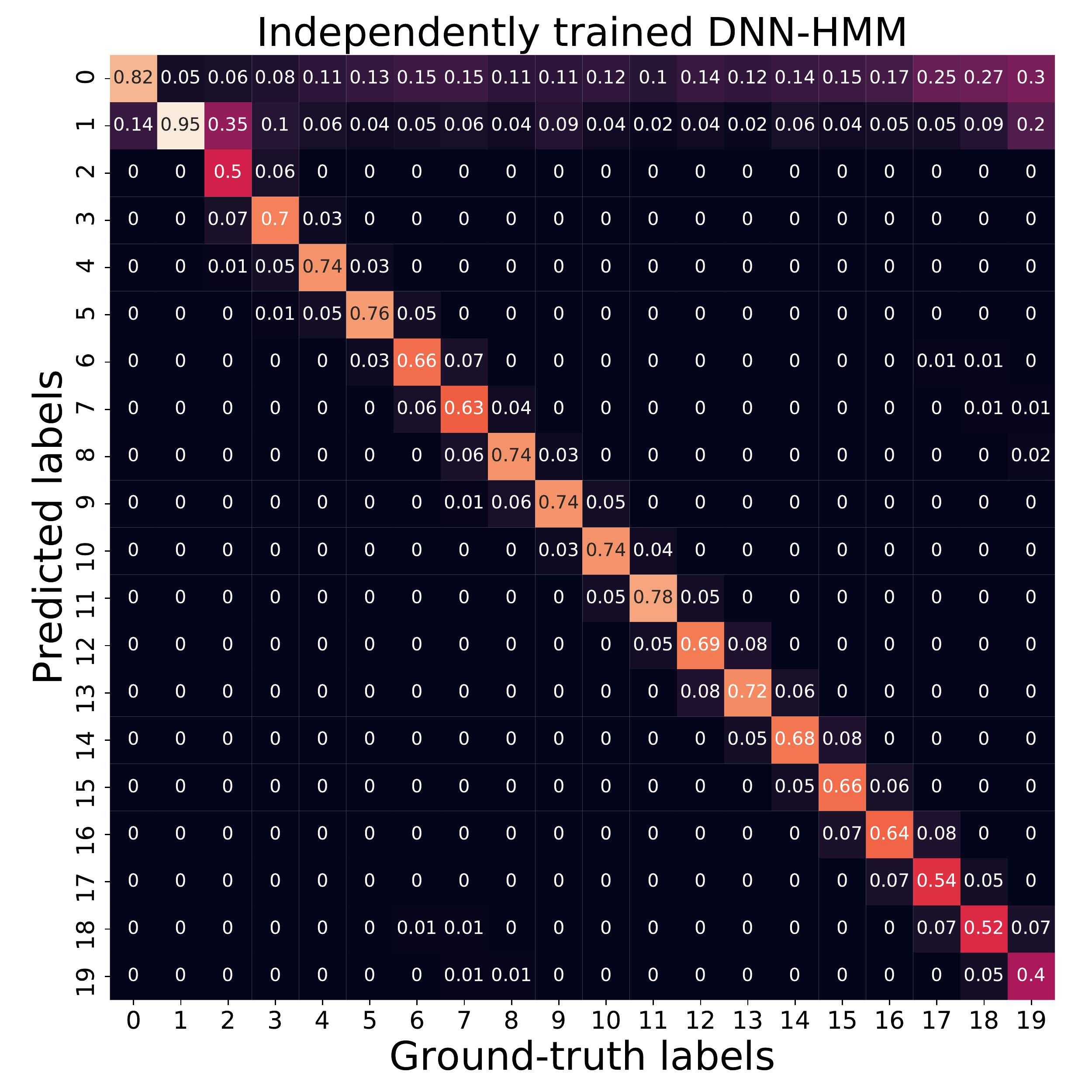}%
}
\\
\vspace{-0.1in}
\subfloat{%
  \includegraphics[clip,width=0.8\columnwidth]{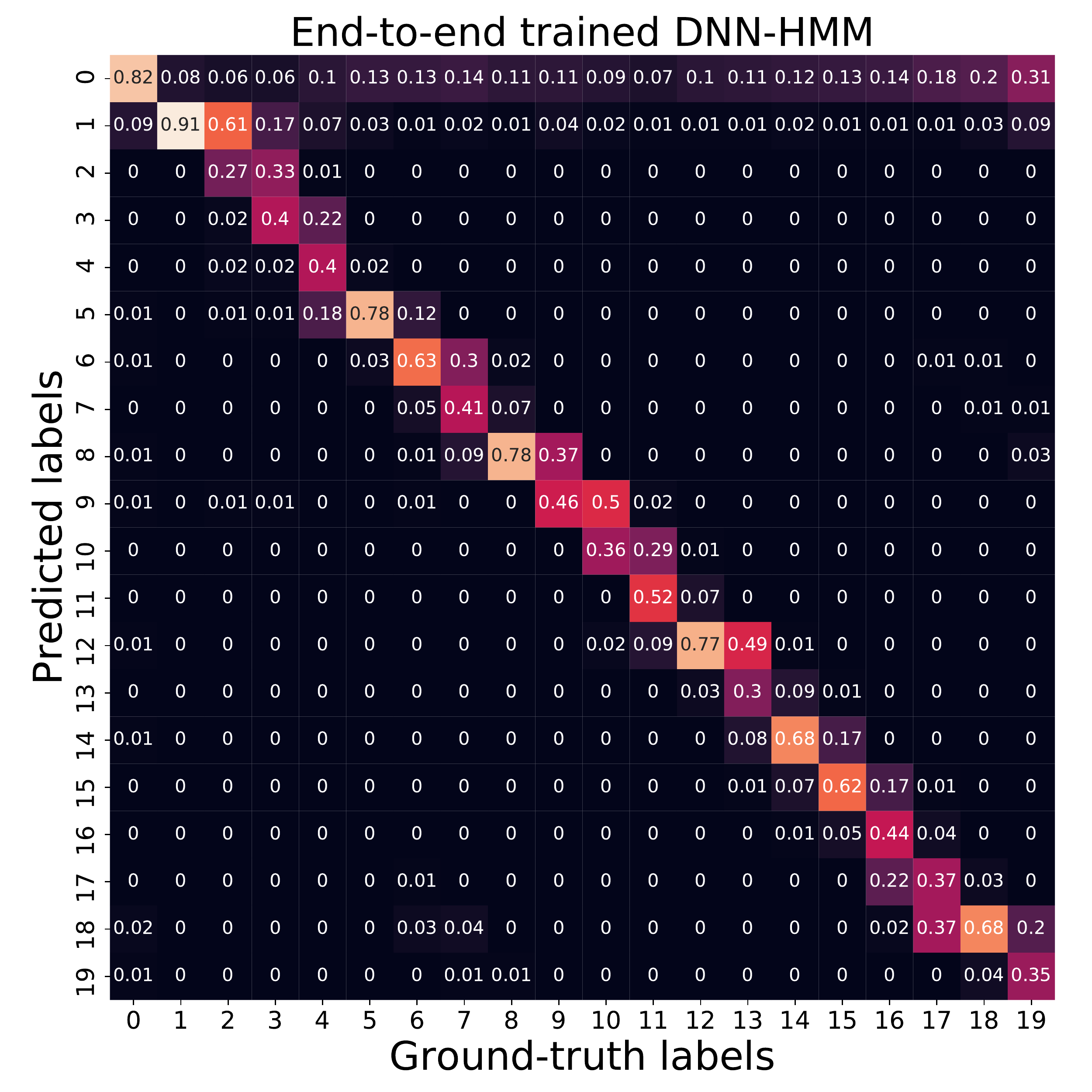}%
}
\vspace{-0.1in}
\caption{Confusion matrices for independently trained DNN-HMM (top) and end-to-end trained DNN-HMM (bottom) models.
The independently trained model spends its capacity in predicting the state probabilities more accurately (average classification accuracy of $68.1\%$), without considering the final detection score.
The end-to-end model addresses this loss-metric mismatch and improves the detection score at the expense of state classification accuracy (of $54.7\%$).
\vspace{-0.2in}}
\label{fig:conf_mtx}
\end{figure}

\begin{figure}[]
    \centering
    \includegraphics[width=.7\linewidth]{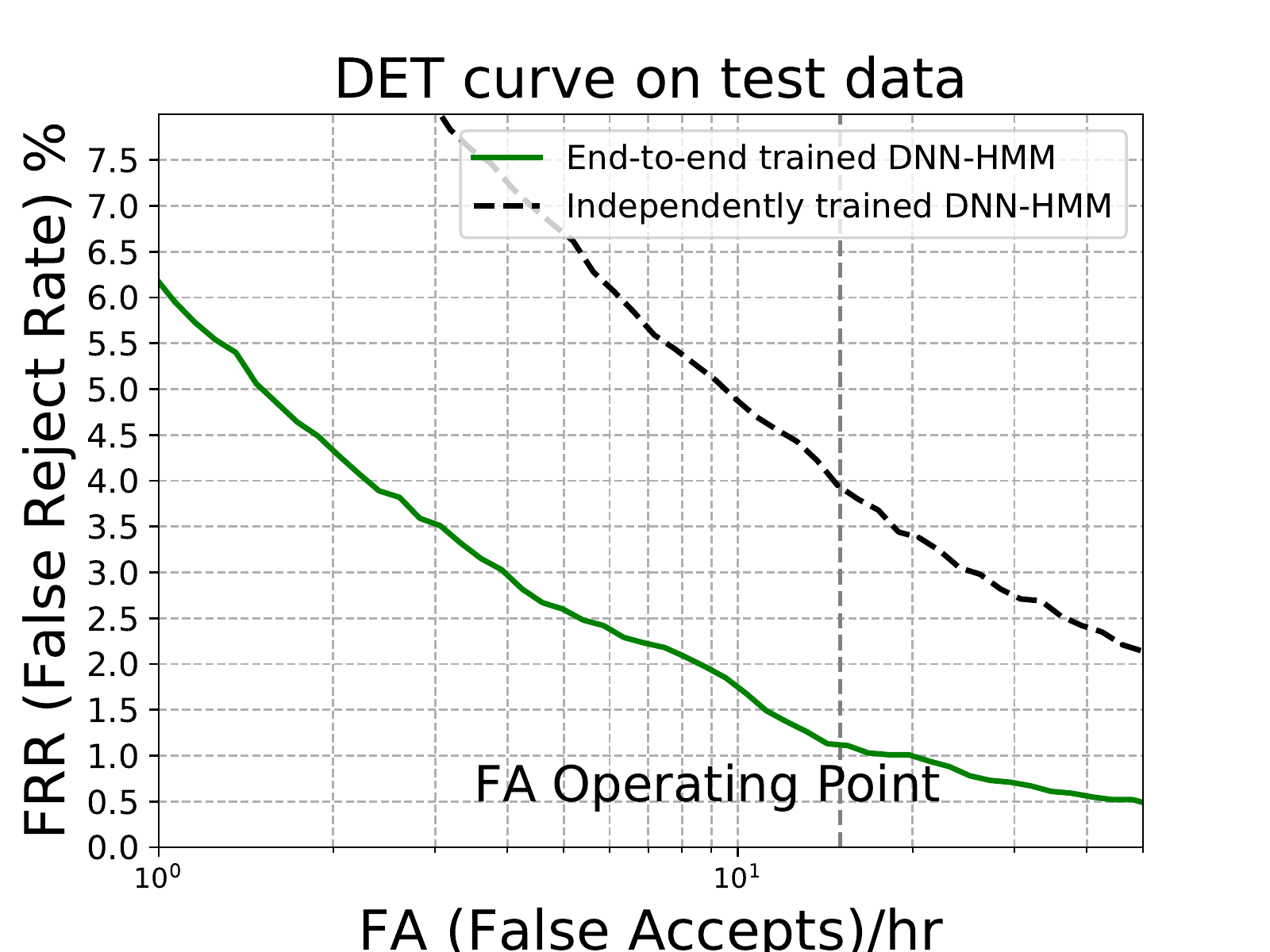} \\
    \vspace{-0.1in}
    \caption{DET-curve for independently trained and end-to-end trained DNN-HMM models on $\approx 2000$ hours of test data.
    The end-to-end trained DNN-HMM has lower FRR at all operating points.
    \vspace{-0.1in}}
    \label{fig:det-curve}
\end{figure}

\subsection{Training and Evaluation}
We first pre-train the DNN model by minimizing the cross-entropy loss in Equation~\eqref{eq:loss_ce}.
Then, we optimize the proposed end-to-end loss in Equation~\eqref{eq:loss_e2e} using Adam optimizer \cite{adam} with an initial learning rate of $0.001$.
The mini-batch consists of positive and negative samples from $48$ random speech utterances, each containing the trigger phrase.
From each training utterance, we select $1$ positive sample with $IOU \ge 0.95$ (i.e.$iou_p = 0.95$) and up to $20$ negative samples with $IOU \le 0.5$ (i.e. $iou_n = 0.5$).
We sample $10$ additional negative windows by splitting the ground-truth window approximately in the middle and swapping their order.
Thus, each mini-batch contains $48$ positive samples and $48 \times 30$ negative samples.
To address the class-imbalance problem, we select $50$ hard negatives (negatives samples with maximum loss value) and $50$ random negatives from all of the negative samples in the mini-batch.

During evaluation, if a detected keyword window overlaps with the ground-truth window, it counts as a true positive (TP), otherwise it counts as a false trigger or false accept (FA).
All the triggers which are not detected are counted as false rejects (FRs).
We vary the detection threshold to compute the detection error tradeoff (DET) curve (FRR vs FA/hr), and compare our method with an independently trained DNN-HMM model in Figure~\ref{fig:det-curve}.
For the proposed method, we note a significant reduction in FRs over all the FA/hr.
We select an operating point at approximately $15$ FA/hr and compute the false reject rate (FRR) in Table~\ref{tab:ModelComparison}.
We also compare our method with a recently proposed Stacked 1D CNN based method~\cite{Higuchi2020} which is roughly at par with our method in FRR.
However, the average $IOU$ (between the ground-truth windows and the predicted true positives windows) of the proposed end-to-end trained DNN-HMM is $27.7\%$ higher than the S1DCNN model, suggesting better localization.
To further measure localization, we compute mean absolute difference of the start and the end locations of the ground-truth windows and the predicted true positives windows.
Our model gives smaller localization error of $0.03$ seconds compared to $0.13$ seconds of S1DCNN.

To further analyze the model, we plot the confusion matrices of the independently trained DNN (that maximize state accuracies) and the end-to-end trained DNN (that optimizes detection score) in Figure~\ref{fig:conf_mtx}.
The DNN optimized for state accuracies distributes its capacity equally to predict the states accurately, leading to higher average state accuracy of $68.1\%$.
However, this does not translate into higher keyword detection accuracy.
In contrast, the end-to-end trained DNN has lower average state accuracy of $54.7\%$ (particularly, there is more confusion among the neighboring states) but focuses on key states that are important for keyword detection, leading to higher keyword detection accuracy.

\begin{table}[!ht]
    \centering
    \begin{tabular}{@{}C{4.5cm}|C{1cm}|C{1.0cm}|C{0.8cm}@{}}
         \toprule[0.015in]
         Model & end-to-end & No. of params   &  FRR ($\%$)\\
         \midrule
         Independently trained DNN-HMM & no &  $13979$  &  $3.95$\\ \midrule
         S1DCNN~\cite{Higuchi2020} & yes & $13993$  &  $1.18$\\ \midrule
         End-to-end trained DNN-HMM & yes & $13979$  &$\mathbf {1.13}$ \\
         \bottomrule
    \end{tabular}
    \vspace{-0.1in}
    \caption{Comparison of FRRs at $15$ FA/hr on test data.
     The end-to-end trained DNN-HMM model has the lowest FRR.}
    \label{tab:ModelComparison}
    \vspace{-0.1in}
\end{table}

%

\section{Conclusion}
In this work, we have addressed the loss-metric mismatch problem of conventional DNN-HMM based keyword detection model by introducing a novel end-to-end training approach using $IOU$ based sampling.
The proposed method works significantly better ($> 70\%$ relative reduction in FRR) than the conventional DNN-HMM training and is more interpretable, accurate in localization, and data-efficient compared to the CNN based end-to-end models.

%




\end{document}